\documentclass[reqno,12pt]{article}
\usepackage{amsmath,amsfonts,amssymb,amsthm,amstext,amscd,eucal,xcolor}
\usepackage[all]{xy}
\usepackage{cite,hyperref}
\usepackage{graphicx}
\usepackage[active]{srcltx}
\makeatletter \@addtoreset{equation}{section}

\makeatletter\renewcommand\section{\@startsection {section}{1}{\z@}%
                                   {-3.5ex \@plus -1ex \@minus -.2ex}
                                   {2.3ex \@plus.2ex}%
                                   {\normalfont\large\bfseries}}
\renewcommand\subsection{\@startsection{subsection}{2}{\z@}%
                                     {-3.25ex\@plus -1ex \@minus -.2ex}%
                                     {1.5ex \@plus .2ex}%
                                     {\normalfont\bfseries}}

\newcommand{\be}{\begin{equation}}
\newcommand{\ee}{\end{equation}}
\newcommand{\bea}{\begin{eqnarray}}
\newcommand{\eea}{\end{eqnarray}}
\newcommand{\bse}{\begin{subequations}}
\newcommand{\ese}{\end{subequations}}
\newcommand{\beqa}{\begin{eqnarray}}
\newcommand{\eeqa}{\end{eqnarray}}
\newcommand{\beqar}{\begin{eqnarray*}}
\newcommand{\eeqar}{\end{eqnarray*}}
\newcommand{\bi}{\begin{itemize}}
\newcommand{\ei}{\end{itemize}}
\newcommand{\bn}{\begin{enumerate}}
\newcommand{\en}{\end{enumerate}}

\newcommand{\ba}{\begin{array}}
\newcommand{\ea}{\end{array}}
\newcommand{\bc}{\begin{center}}
\newcommand{\ec}{\end{center}}

\definecolor{darkgreen}{rgb}{0,0.3,0}
\definecolor{darkblue}{rgb}{0,0,0.3}
\definecolor{darkred}{rgb}{0.7,0,0}

\begin{document}
\begin{titlepage}
\vspace{0.5cm}
\begin{center}
\centerline{{\bf{Higher derivative Chern-Simons extension in the noncommutative QED$_{3}$}}} \vspace{6mm}
\bf{M. Ghasemkhani\footnote{e-mail:  $m_{_{-}}$ghasemkhani@sbu.ac.ir}$^{;\ a}$
 and R. Bufalo\footnote{e-mail: rodrigo.bufalo@helsinki.fi}$^{;\ b,c}$ }
\\
\vspace{5mm}
\normalsize
\bigskip\medskip
{$^a$ \it  Department of Physics, Shahid Beheshti University, \\G.C., Evin, Tehran 19839, Iran}\\
{$^b$ \it Department of Physics, University of Helsinki, P.O. Box 64\\ FI-00014 Helsinki, Finland}\\
\smallskip
{$^c$ \it Instituto de F\'{\i}sica Te\'orica (IFT), Universidade Estadual Paulista\\
Rua Dr. Bento Teobaldo Ferraz 271, Bloco II, 01140-070 S\~ao Paulo, SP, Brazil}\\
 \end{center}
\begin{abstract}
\noindent The noncommutative (NC) massive quantum electrodynamics in $2+1$ dimensions is considered. We show explicitly that the one-loop effective action arising from the integrating out the fermionic fields leads to the ordinary NC Chern-Simons and NC Maxwell action at the long wavelength limit (large fermion mass). In the next to
leading order, the higher-derivative contributions to NC Chern-Simons are obtained. Moreover, the gauge invariance of the outcome action is carefully discussed. We then consider the higher-derivative modification into the pure NC Chern-Simons Lagrangian density and evaluate the one-loop correction to the pole of the photon propagator.

\end{abstract}
\end{titlepage}
\setcounter{footnote}{0}
\renewcommand{\baselinestretch}{1.05}  


\tableofcontents
\section{Introduction}\label{sec-intro}
Low-dimensional field theories were recognized, a long time ago, as serving as laboratories where important theoretical ideas are suitably tested in a simple setting, specially on condensed matter systems. Besides, these studies were also driven with the wishful thought that in such simpler setting we can learn useful things about the well-recognized four-dimensional problems.
Furthermore, including to that, the fact that a field theory defined in a three-dimensional space-time contains a highly interesting inner structure, due to the odd space-time dimensionality, it is rather natural to investigate them as theoretical options \cite{ref7,ref8,ref12,ref15}.

Due to the recent improvement concerning the precision of the measurements of experiments (LHC, ILC, etc) investigating particles properties, the possibility of observing direct evidence of new physics has captured and led to a major interest in the development and understanding of the physics in higher scales, for instance at Planck scale. A conscientious point in the majority of the analyses is that the nature of the space-time may change at Planck scale. This fact had strong influence in thinking that the physics at Planck scale may suitably be described by a noncommutativity between the coordinates of the space-time. Direct motivations supporting considering models constructed on noncommutative space-time are many, and come from several theoretical areas such as string theory, quantum gravity, and Lorentz breaking  \cite{ref9,ref29,ref30}.

In recent years, we have witnessed an enormous activity involving the development of noncommutative (NC) gauge theories in high-energy physics \cite{ref9,ref10}. The noncommutativity of space-time, whose structure is determined by $[x^\mu,x^\nu]=i\theta^{\mu\nu}$, has provided a better understanding about the quantum nature of space-time. On the other hand, it has been also studied how noncommutativity affects established properties of conventional theories, i.e., studying NC extensions of well studied quantum field theories and to look then for NC effects on its deviations, since it is generally found that such extensions behave in a very interesting and nontrivial ways. Despite its close relation with the space-time structure, no consistent gravitational theory defined on noncommutative space-time has been established yet \cite{ref19}.

In fact, various studies in analyzing three-dimensional gauge theories defined in a noncommutative space-time have uncovered deviations of known phenomena and interesting new properties of these theories \cite{ref16,ref17,ref1,ref5,ref6,ref13,ref18}. One of the most exploited features of a three-dimensional gauge theory is that, in one side, if you started with a theory with massless fermionic fields interacting with Chern-Simons gauge fields, a mass for the fermionic fields is generated dynamically by radiative corrections; on the other side, if you started with a theory with massive fermionic fields interacting with an external gauge field, the Chern-Simons action is also induced by radiative corrections. These properties are usually related to nonperturbative phenomena due to the space-time dimensionality.

The generation of the Chern-Simons action in a noncommutative gauge theory was the subject of several analyses, with the most different purposes \cite{ref6,ref18,ref21,ref22,ref23,ref24,ref25}. However, no higher-derivative (HD) extensions of the Chern-Simons action have been considered. This subject was first discussed in great detail by considering the full Abelian action, consisting in the Chern-Simons added by the Maxwell theory, \footnote{This theory is the so-called topologically massive electrodynamics \cite{ref7}, and describes a helicity $\pm 1$ mode.}, adjoined by higher-derivative contributions appearing from a $( \partial / m)$ perturbative expansion of the effective action of QED$_3$ \cite{ref2,ref3}. Our main aim in the present paper is to treat the latter phenomenon in full detail when formulated within the framework of noncommutative gauge theory.

In this paper, we discuss, within the effective action approach, the one-loop properties of the three-dimensional noncommutative QED. For this purpose, we follow the idea outlined in Ref.\cite{ref11} in which noncommutative fermionic effective actions were considered. Although the Seiberg-Witten map is perturbative by nature \cite{ref9}, remarkably, it suffices to consider the existence of an exact Seiberg-Witten map, valid to all orders in $\theta$, in a formal way that the noncommutative effects into the resulting outcome are in fact nonperturbative. For this purpose, in Sec.~\ref{sec:2}, we review the basic ideas consisting in this nonperturbative approach (in $\theta$) to the NC QED$_3$, obtaining the main objects of our analysis at one-loop effective action. Furthermore, in Sec.~\ref{sec:3}, we perform explicitly the calculation for two, three and four gauge fields of the one-loop effective action. In particular, we will consider the long wavelength limit (large fermion mass) and consider the terms in the expansion $\mathcal{O}(m^ 0)$, $\mathcal{O}(m^ {-1})$ and $\mathcal{O}(m^{-2})$, which correspond exactly to the NC Chern-Simons action, NC Maxwell action, and higher-derivative extension to the NC Chern-Simons action, respectively. Also, the gauge invariance of the higher-derivative extension is discussed.
By completion we present a discussion to characterize the excitation modes regarding the effective action of the HD extension of the pure NC Chern-Simons theory in Sec.~\ref{sec:4}. As well as the one-loop correction, arising from the higher-derivative terms, to the pole of the photon propagator in the infrared limit is studied, which leads to the noncommutative UV/IR mixing effect. In Sec.~\ref{sec:5} we summarize the results, and present our final remarks.
\section{General discussion}
\label{sec:2}
In this section, we will introduce our basic notation and describe the analysis method. Let us consider the noncommutative extension of fermionic fields interacting
in the presence of an external gauge field. For this, we shall consider the following action:
\begin{equation}
S=\int d^{3}x\left[\bar\psi\star i\gamma^{\mu}D_{\mu}^{\star}\psi-m\bar\psi\star\psi\right],\label{eq: 0.1}
\end{equation}
in which the covariant derivative is defined as $D_{\mu}^{\star}\psi=\partial _{\mu} \psi-igA_{\mu}\star\psi$. The action is invariant under the infinitesimal gauge transformations,
\begin{equation}
\delta A _\mu  = \partial _\mu \lambda - ig \left[A_\mu , \lambda\right]_\star,\quad \delta \psi=i g \lambda \star \psi.
\end{equation}
Moreover, it should be emphasized that we are working with a two-component representation for the fermionic fields. In this representation, the $\gamma$-matrices satisfy $\gamma ^\mu \gamma ^\nu = \eta ^{\mu \nu}-i \epsilon^{\mu\nu \alpha}\gamma_\alpha $. Furthermore, we introduce the Moyal star product between the functions $f$ and $g$ defined as
\begin{equation}
f\left(x\right)\star g\left(x\right)=f\left(x\right)\exp\bigg(\frac{i}{2}
\theta ^{\mu\nu} \overleftarrow{\partial_{\mu}}~\overrightarrow{\partial_{\nu}}\bigg)g\left(x\right), \label{eq: 0.1a}
\end{equation}
where we assume that the noncommutative structure of the space-time is determined by $[x^\mu,x^\nu]=i\theta ^{\mu\nu}$, in which $\theta^{\mu\nu}=-\theta^{\nu\mu}$ are constant parameters.
The one-loop effective action can be readily obtained by integrating out the fermionic fields of \eqref{eq: 0.1},
\begin{equation}
i\Gamma\left[A\right]=\ln\frac{\det\left(\displaystyle{\not}D^{\star}+im\right)}
{\det\left(\displaystyle{\not}\partial+im\right)}
=-\sum_{n}\frac{1}{n}tr\bigg[\left(\displaystyle{\not}\partial+im\right)^{-1}ig \displaystyle{\not}A\star\bigg]^{n}.\label{eq: 0.2}
\end{equation}
The differential operator in \eqref{eq: 0.2} is identified as being the fermionic propagator,
\begin{equation}
\left(\displaystyle{\not}\partial+im\right)^{-1}\delta\left(x-y\right)=
\int\frac{d^{3}p}{\left(2\pi\right)^{3}}
\frac{i\left(\displaystyle{\not}p+m\right)}{p^{2}-m^{2}+i\varepsilon}e^{-ip.\left(x-y\right)}.
\end{equation}
Nevertheless, we can rewrite \eqref{eq: 0.2} in a more suitable form as for perturbative computation,
\begin{equation}
i\Gamma\left[A\right]=\sum_{n}\int d^{3}x_{1}\ldots\int d^{3}x_{n}A_{\mu_{1}}\left(x_{1}\right)A_{\mu_{2}}\left(x_{2}\right)\cdots A_{\mu_{n}}\left(x_{n}\right)\Gamma^{\mu_{1}\mu_{2}\ldots\mu_{n}}\left(x_{1},x_{2},\ldots,x_{n}\right),\label{eq: 0.3}
\end{equation}
where we have written the one-loop contributions from the noncommutative gauge field such as
\begin{align}
\Gamma^{\mu_{1}\mu_{2}\ldots\mu_{n}}\left(x_{1},x_{2},\ldots x_{n}\right)=&-\frac{\left(-g\right)^{n}}{n}\int\prod_{i}\frac{d^{3}p_{i}}{\left(2\pi\right)^{3}}
\left(2\pi\right)^{3}\delta(\sum_{i}p_{i}) \exp(-i\sum_{i}p_{i}.x_{i})\nonumber\\
 &\times   \exp\left(-\frac{i}{2}\sum_{i<j}p_{i}\times p_{j}\right)\Xi^{\mu_{1}\mu_{2}\ldots\mu_{n}}\left(p_{1},p_{2},\ldots,p_{n-1}\right),\label{eq: 0.4}
\end{align}
in which we have introduced the notation $p\times q=\theta^{\mu\nu}p_{\mu}q_{\nu}$, and by simplicity we have also defined the one-loop contributions in the form,
\begin{align}
\Xi^{\mu_{1}\mu_{2}\ldots\mu_{n}}=
\int\frac{d^{3}q}{\left(2\pi\right)^{3}}
\frac{tr\bigg[\left(\displaystyle{\not}q+\displaystyle{\not}p_{1}
+m\right)
\gamma^{\mu_{1}}\left(\displaystyle{\not}q+m\right)
\gamma^{\mu_{2}}\left(\displaystyle{\not}q-\displaystyle{\not}p_{2}+m\right)
\gamma^{\mu_{3}}\ldots(\displaystyle{\not}q-
\sum\limits_{i=2}^{n-1}\displaystyle{\not}p_{i}+m)\gamma^{\mu_{n}}\bigg]}
{\left[\left(q+p_{1}\right)
^{2}-m^{2}\right]\left[q^{2}-m^{2}\right]\left[\left(q-p_{2}\right)^{2}-m^{2}\right]
\ldots\left[(q-\sum\limits_{i=2}^{n-1}p_{i})^{2}-m^{2}\right]}
.\label{eq: 0.5}
\end{align}
It should be stressed that in order to rewrite \eqref{eq: 0.2} into the form \eqref{eq: 0.3}, we have made use of the general result,
\begin{align}
\int d^{3}x\left[\mathcal{O}_{1}\left(x\right)\star\mathcal{O}_{2}\left(x\right)
\ldots\star\mathcal{O}_{n}\left(x\right)\right]&=\int\prod_{i}d^{3}x_{i}
\prod_{i}\frac{d^{3}p_{i}}{\left(2\pi\right)^{3}}~[\mathcal{O}_{1}\left(x_{1}\right)\mathcal{O}_{2}\left(x_{2}\right)
\ldots\mathcal{O}_{n}\left(x_{m}\right)]\notag \\
&\times \exp(-i\sum_{i}p_{i}.x_{i})\exp\left(-\frac{i}{2}\sum_{i<j}p_{i}\times p_{j}\right)\delta(\sum_{i}p_{i}).\label{eq: 0.6}
\end{align}
Now that we have concluded with our formal development and presented all the necessary information, we will proceed in evaluating explicitly the contributions for $n=2,3,4$ gauge fields
in \eqref{eq: 0.3}. Initially we will consider and evaluate the general expressions for such contributions; however, by means of illustration we will consider in particular the resulting expressions in the long wavelength limit (i.e., $m^2 \gg p^2$, where $p$ is an external momenta). Moreover, we will concentrate on the leading contributions $\mathcal {O}\left(m^0\right)$, $\mathcal {O}\left(m^{-1}\right)$, and $\mathcal {O}\left(m^{-2}\right)$, which are responsible to generate the noncommutative Chern-Simons, Maxwell, and HD Chern-Simons extension, respectively.

\section{Perturbative effective action}
\label{sec:3}
We shall now proceed in evaluating explicitly the one-loop contributions of \eqref{eq: 0.3} for two, three and four gauge fields. Actually, the contribution of one gauge field of \eqref{eq: 0.3} is identically vanishing. Among all the terms from such contributions, we will show that at the long wavelength limit the complete three-dimensional action for the noncommutative Chern-Simons and Maxwell theory is generated; moreover, we also discuss the higher-derivatives extension for the noncommutative Chern-Simons action. \footnote{Indeed, a complete discussion of all higher-derivative contributions at order $sgn(m)/m^2$ is presented (and generalized) in the Appendix \ref{sec:appA}.}

\begin{figure}[h]
\vspace{-2.5cm}
 \includegraphics[width=13cm,height=8cm]{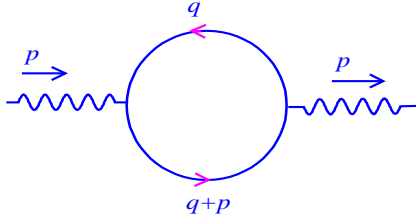}
   \vspace{-2.5cm}\centering
  \caption{\small{Relevant graph for the AA-term.}
}\label{AA-graph}
\end{figure}
\subsection{AA-term contribution}
In order to evaluate the first nonvanishing contribution, let us take $n=2$ in the one-loop contribution \eqref{eq: 0.5}, which is shown in Fig.~\ref{AA-graph}:
\begin{equation}
\Xi^{\mu\nu}\left(p\right)=\int\frac{d^{3}q}{\left(2\pi\right)^{3}}\frac{tr\left[\left(\displaystyle{\not}q+\displaystyle{\not}p+m\right)\gamma^{\mu}\left(\displaystyle{\not}q+m\right)\gamma^{\nu}\right]}{\left[\left(q+p\right)^{2}-m^{2}\right]\left[q^{2}-m^{2}\right]}.\label{eq: 0.9}
\end{equation}
This expression can be put into a more convenient form, which facilitates the evaluation of the momentum integration. Thus, we can make use of the Feynman parametrization, which combined
with the change of variables $q\rightarrow q+xp$ results into
\begin{equation}
\Xi^{\mu\nu}\left(p\right)=\int_{0}^{1}dx\int\frac{d^{3}q}{\left(2\pi\right)^{3}}
\frac{tr\left[\left(\displaystyle{\not}q+\left(1-x\right)\displaystyle{\not}p+m\right)
\gamma^{\mu}\left(\displaystyle{\not}q-x\displaystyle{\not}p+m\right)\gamma^{\nu}\right]}
{\left[q^{2}+x\left(1-x\right)p^{2}-m^{2}\right]^{2}}.\label{eq: 0.10}
\end{equation}
We can evaluate the trace in \eqref{eq: 0.10} and write the integration into a dimensional regularization form. Hence, with the known results for a two-dimensional representation,
\begin{equation}
tr\left(\gamma^{\mu}\gamma^{\nu}\right)=2\eta^{\mu\nu},\quad tr\left(\gamma^{\mu}\gamma^{\beta}\gamma^{\nu}\right)=2i\epsilon^{\mu\beta\nu},\quad tr\left(\gamma^{\alpha}\gamma^{\mu}\gamma^{\beta}\gamma^{\nu}\right)=2\left(\eta^{\alpha\mu}\eta^{\beta\nu}-\eta^{\alpha\beta}\eta^{\mu\nu}+\eta^{\alpha\nu}\eta^{\mu\beta}\right), \label{eq: 0.11}
\end{equation}
we obtain \cite{ref61}
\begin{equation}
\Xi^{\mu\nu}\left(p\right)=2\int_{0}^{1}dx\int\frac{d^{d}q}{\left(2\pi\right)
^{d}}\frac{\left(\frac{2}{d}-1\right)q^{2}\eta^{\mu\nu}-2x\left(1-x\right)p^{\mu}p^{\nu}+imp_{\alpha}\epsilon^{\alpha\mu\nu}+\left(m^{2}+x\left(1-x\right)p^{2}\right)\eta^{\mu\nu}}{\left[q^{2}+x\left(1-x\right)p^{2}-m^{2}\right]^{2}}.
\end{equation}
Here $d$ denotes the dimension of the space-time. The detailed evaluation of the momentum integral is rather direct, taking the limit $d\rightarrow 3$
and the resulting expression reads
\begin{align}
\Xi^{\mu\nu}\left(p\right) & =-\frac{1}{4\pi}mp_{\alpha}\epsilon^{\alpha\mu\nu}
\int_{0}^{1}dx\frac{1}{\left(m^{2}-x\left(1-x\right)p^{2}\right)^{\frac{1}{2}}}+
\frac{i}{2\pi}\left(p^{2}\eta^{\mu\nu}-p^{\mu}p^{\nu}\right)\int_{0}^{1}
dx\frac{x\left(1-x\right)}{\left(m^{2}-x\left(1-x\right)p^{2}\right)^{\frac{1}{2}}}.\label{eq: 0.12}
\end{align}
\subsubsection{Higher derivative contribution}
Let us take a closer look at the expression \eqref{eq: 0.12}. The higher-derivative contributions are obtained by considering the long wavelength limit $p^{2}\ll m^{2}$. Thus expanding the expression and performing the remaining integrals, we obtain
\begin{align}
\Xi^{\mu\nu}\left(p\right) & =-\frac{sgn\left(m\right)}{4\pi}\left(p_{\alpha}\epsilon^{\alpha\mu\nu}\right)
\left(1+\frac{p^{2}}{12m^{2}}\right)+\frac{i}{12\pi}\frac{1}{\left|m\right|}
\left(p^{2}\eta^{\mu\nu}-p^{\mu}p^{\nu}\right)+\mathcal{O}\left(\frac{p^{4}}{m^{4}}\right),\label{eq: 0.13}
\end{align}
 in which, $sgn\left(m\right)\equiv \frac{m}{\left|m\right|}$ is the sign function. As it is easily seen, the first term corresponds to the kinetic term of the
Chern-Simons term, $\mathcal {O}\left(m^{0}\right)$, and its HD extension, $\mathcal {O}\left(m^{-2}\right)$, while the second one is the kinetic part of the noncommutative Maxwell action, $\mathcal {O}\left(m^{-1}\right)$. By means of illustration, let us consider the first term from \eqref{eq: 0.13} which corresponds to the Chern-Simons contribution
\begin{align}
\Xi_{\tiny\mbox{CS}}^{\mu\nu}\left(p\right) & =-\frac{sgn\left(m\right)}{4\pi}\left(p_{\alpha}\epsilon^{\alpha\mu\nu}\right)\left(1+\frac{p^{2}}{12m^{2}}+\mathcal{O}\left(\frac{p^{4}}{m^{4}}\right)\right). \label{eq: 0.13a}
\end{align}
Substituting it into the expression \eqref{eq: 0.4} and after some integral manipulation, we find
\begin{align}
\Gamma_{\tiny\mbox{CS}}^{\mu\nu}\left(x_{1},x_{2}\right) & =-\frac{g^{2}}{2}\int\frac{d^{3}p}{\left(2\pi\right)^{3}}\exp\left[-ip.\left(x_{1}-x_{2}\right)\right]\Xi_{cs}^{\mu\nu}\left(p\right), \nonumber \\
 & =-i\frac{g^{2}}{8\pi}sgn\left(m\right)\epsilon^{\alpha\mu\nu}
 \left(1-\frac{\square}{12m^{2}}\right)\partial_{\alpha}\delta\left(x_{1}-x_{2}\right). \label{eq: 0.14}
\end{align}
 Finally, we substitute \eqref{eq: 0.14} into \eqref{eq: 0.3} to then obtain the expression for the Chern-Simons effective action
\begin{align}
i\Gamma_{\tiny\mbox{CS}}\left[AA\right] & =i\frac{g^{2}}{8\pi}sgn\left(m\right)\epsilon^{\mu\alpha\nu}\int d^{3}x\left[A_{\mu}\left(x\right)\left(1-\frac{\square}{12m^{2}}\right)\partial_{\alpha}A_{\nu}\left(x\right)\right]. \label{eq: 0.15}
\end{align}
 As it is well known, the expression for two gauge fields \eqref{eq: 0.15} does not display effects from noncommutativity, since the phase factor from \eqref{eq: 0.4} vanishes for $n=2$.
Furthermore, one can perform the same manipulation as above and realize that the second term in \eqref{eq: 0.13}, when written in terms of
the effective action, is the quadratic part of the noncommutative Maxwell action
\begin{align}
i\Gamma_{\tiny\mbox{M}}\left[AA\right] & =i\frac{g^{2}}{24 \pi}\frac{1}{\left|m\right|} \int d^{3}x\left[\partial^{\nu} A_{\mu}\partial_{\nu}A^{\mu}
-\partial^{\nu} A_{\mu}\partial^{\mu}A_{\nu}\right]. \label{eq: 0.16}
\end{align}
\subsection{AAA-term contribution}
\begin{figure}[h]
\vspace{-2.7cm}
  \includegraphics[width=12cm,height=8cm]{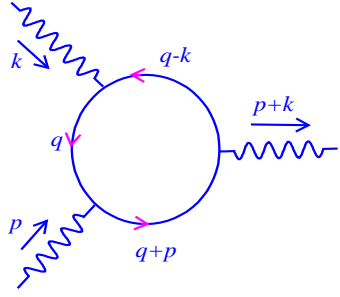}
   \vspace{-2cm}\centering
  \caption{Relevant graph for the AAA-term.
}\label{AAA-graph}
\end{figure}

In the same way as the previous calculation, we start by evaluating the one-loop contribution \eqref{eq: 0.5} for the case $n=3$, represented in Fig.~\ref{AAA-graph}:
\begin{equation}
\Xi^{\mu\nu\sigma}\left(p,k\right)=\int\frac{d^{3}q}{\left(2\pi\right)^{3}}\frac{tr\left[\left(\displaystyle{\not}q+\displaystyle{\not}p+m\right)\gamma^{\mu}\left(\displaystyle{\not}q+m\right)\gamma^{\nu}\left(\displaystyle{\not}q-\displaystyle{\not}k+m\right)\gamma^{\sigma}\right]}{\left[\left(q+p\right)^{2}-m^{2}\right]\left[q^{2}-m^{2}\right]\left[\left(q-k\right)^{2}-m^{2}\right]}.\label{eq: 1.1}
\end{equation}
Moreover, we can use the Feynman parametrization and the change of variables $q\rightarrow q-\left(xp-zk\right)\equiv q-s$ to obtain
\begin{equation}
\Xi^{\mu\nu\sigma}\left(p,k\right)=2\int d\xi\int\frac{d^{3}q}{\left(2\pi\right)^{3}}\frac{tr\left[\left(\displaystyle{\not}q-\displaystyle{\not}s+\displaystyle{\not}p+m\right)\gamma^{\mu}\left(\displaystyle{\not}q-\displaystyle{\not}s+m\right)\gamma^{\nu}\left(\displaystyle{\not}q-\displaystyle{\not}s
-\displaystyle{\not}k+m\right)\gamma^{\sigma}\right]}{\left[q^{2}+A^{2}\left(p,k\right)-m^{2}\right]^{3}},\label{eq: 1.3}
\end{equation}
\vspace{-0.5cm}

\noindent
where we have introduced the notation $\int d\xi=\int_{0}^{1}dx\int_{0}^{1-x}dz$, and $A^{2}\left(p,k\right)=-\left(xp-zk\right)^{2}+xp^{2}+zk^{2}$. The momentum integration can be performed without any complication by using standard methods. Hence, after separating the different powers of $q$ in the numerator and then evaluating the integration, we find
\begin{align}
\Xi^{\mu\nu\sigma}\left(p,k\right) & =\frac{i}{16\pi}\int d\xi\left[\frac{A^{\mu\nu\sigma}\left(p,k;x,z\right)}{\left(m^{2}-A^{2}\left(p,k\right)\right)^{\frac{1}{2}}}-\frac{B^{\mu\nu\sigma}\left(p,k;x,z\right)}{\left(m^{2}-A^{2}\left(p,k\right)\right)^{\frac{3}{2}}}\right],\label{eq: 1.5}
\end{align}
by sake of notation, we have introduced the quantities, with $s_{\mu}=xp_{\mu}-zk_{\mu}$,
\begin{align}
A^{\mu\nu\sigma}\left(p,k;x,z\right)  =&tr\left[\gamma^{\alpha}\gamma^{\mu}\gamma_{\alpha}\gamma^{\nu}\left(-\displaystyle{\not}s-\displaystyle{\not}k+m\right)
\gamma^{\sigma}\right]
-tr\left[\gamma^{\alpha}\gamma^{\mu}\displaystyle{\not}s\gamma^{\nu}\gamma_{\alpha}\gamma^{\sigma}\right] \notag \\
 & +mtr\left[\gamma^{\alpha}\gamma^{\mu}\gamma^{\nu}\gamma_{\alpha}\gamma^{\sigma}\right]
 -tr\left[\displaystyle{\not}s\gamma^{\mu}\gamma^{\alpha}\gamma^{\nu}\gamma_{\alpha}\gamma^{\sigma}\right]\notag \\
 &+tr\left[\displaystyle{\not}p\gamma^{\mu}\gamma^{\alpha}\gamma^{\nu}\gamma_{\alpha}\gamma^{\sigma}\right]
 +mtr\left[\gamma^{\mu}\gamma^{\alpha}\gamma^{\nu}\gamma_{\alpha}\gamma^{\sigma}\right], \label{eq: 1.6}
\end{align}
and
\begin{align}
B^{\mu\nu\sigma}\left(p,k;x,z\right)  =& tr\left[\displaystyle{\not}s\gamma^{\mu}\displaystyle{\not}s\gamma^{\nu}
\left(-\displaystyle{\not}s-\displaystyle{\not}k+m\right)\gamma^{\sigma}\right]-mtr\left[\displaystyle{\not}s\gamma^{\mu}\gamma^{\nu}\left(-\displaystyle{\not}s-\displaystyle{\not}k+m\right)\gamma^{\sigma}\right]\nonumber \\
 & -tr\left[\displaystyle{\not}p\gamma^{\mu}\displaystyle{\not}s\gamma^{\nu}\left(-\displaystyle{\not}s-\displaystyle{\not}k+m\right)\gamma^{\sigma}
 \right]+mtr\left[\displaystyle{\not}p\gamma^{\mu}\gamma^{\nu}\left(-\displaystyle{\not}s-\displaystyle{\not}k+m\right)\gamma^{\sigma}\right]\nonumber \\
 & -mtr\left[\gamma^{\mu}\displaystyle{\not}s\gamma^{\nu}\left(-\displaystyle{\not}s-\displaystyle{\not}k+m\right)\gamma^{\sigma}\right]+m^{2}tr\left[\gamma^{\mu}\gamma^{\nu}
 \left(-\displaystyle{\not}s-\displaystyle{\not}k+m\right)\gamma^{\sigma}\right]. \label{eq: 1.7}
\end{align}
 As usual, we could analyze in detail the full contribution of \eqref{eq: 1.5}, but since we are interested in particular cases that induce the $AAA$ noncommutative parts of the Chern-Simons and Maxwell action, we will retain our attention into the $\mathcal{O}\left(m^{0}\right)$, $\mathcal{O}\left(m^{-1}\right)$ and $\mathcal{O}\left(m^{-2}\right)$ contributions.

\subsubsection{Higher derivative contribution}
The contributions for the effective action from $AAA$ gauge fields are more complicated than those from the $AA$ fields; this is due to the structure of the denominator and numerator
of the terms from the expression (\ref{eq: 1.5}). Nevertheless, the contributions for the Chern-Simons action and its HD extension are those associated with terms of order $\mathcal{O}\left(m^{0}\right)$ and $\mathcal{O}\left(m^{-2}\right)$. Hence, by simple power counting, we consider those terms that are linear in $m$ from \eqref{eq: 1.6},
\begin{align}
A_{\tiny\mbox{CS}}^{\mu\nu\sigma} & \left(p,k;x,z\right)= - mtr\left[\gamma^{\mu}\gamma^{\nu}\gamma^{\sigma}\right]-m
tr\left[\gamma^{\mu}\gamma^{\nu}\gamma^{\sigma}\right] -mtr\left[\gamma^{\mu}\gamma^{\nu}\gamma^{\sigma}\right], \label{eq: 1.8}
\end{align}
and those that are linear and cubic in $m$ from \eqref{eq: 1.7},
\begin{align}
B_{\tiny\mbox{CS}}^{\mu\nu\sigma}\left(p,k;x,z\right) & =mtr\left[\left(\displaystyle{\not}s-\displaystyle{\not}p\right) \gamma^{\mu}\displaystyle{\not}s\gamma^{\nu}\gamma^{\sigma}\right]+mtr\left[\left(\displaystyle{\not}s-\displaystyle{\not}p\right)\gamma^{\mu}\gamma^{\nu}
\left(\displaystyle{\not}s+\displaystyle{\not}k\right)\gamma^{\sigma}\right] \nonumber \\
 &+mtr\left[\gamma^{\mu}\displaystyle{\not}s\gamma^{\nu}
 \left(\displaystyle{\not}s+\displaystyle{\not}k\right)\gamma^{\sigma}\right]
 +m^{3}tr\left[\gamma^{\mu}\gamma^{\nu}\gamma^{\sigma}\right], \label{eq: 1.9}
\end{align}
while, for the Maxwell action we shall consider the terms $\mathcal{O}\left(m^{-1}\right)$, which are those terms independent of $m$ in \eqref{eq: 1.6},
\begin{align}
A_{\tiny\mbox{M}}^{\mu\nu\sigma}\left(p,k;x,z\right) & = tr\left[\gamma^{\mu} \gamma^{\nu}\left(\displaystyle{\not}s+\displaystyle{\not}k\right)\gamma^{\sigma}\right]+
tr\left[\gamma^{\mu}\displaystyle{\not}s\gamma^{\nu}\gamma^{\sigma}\right]  + tr\left[\displaystyle{\not}s\gamma^{\mu} \gamma^{\nu} \gamma^{\sigma}\right]-
 tr\left[\displaystyle{\not}p\gamma^{\mu} \gamma^{\nu}\gamma^{\sigma}\right], \label{eq: 1.10}
\end{align}
and quadratic in $m$ from \eqref{eq: 1.7},
\begin{align}
B_{\tiny\mbox{M}}^{\mu\nu\sigma}\left(p,k;x,z\right) & =-m^{2}tr\left[\displaystyle{\not}s\gamma^{\mu}\gamma^{\nu}\gamma^{\sigma}\right]+m^{2}tr\left[\displaystyle{\not}p\gamma^{\mu}\gamma^{\nu}\gamma^{\sigma}\right]\notag \\
&-m^{2}tr\left[\gamma^{\mu}\displaystyle{\not}s\gamma^{\nu}\gamma^{\sigma}\right]-m^{2}tr\left[\gamma^{\mu}\gamma^{\nu}\left(\displaystyle{\not}s+\displaystyle{\not}k\right)\gamma^{\sigma}\right]. \label{eq: 1.11}
\end{align}
Let us now focus our attention into the Chern-Simons contributions. We can simplify the above equations by means of the Clifford algebra $\left\{ \gamma^{\mu},\gamma^{\nu}\right\} =2\eta^{\mu\nu}$ and the identities \eqref{eq: 0.11}, resulting in the following expressions
\begin{align}
A_{\tiny\mbox{CS}}^{\mu\nu\sigma}\left(p,k;x,z\right) & =-6im\epsilon^{\mu\nu\sigma},\label{eq: 1.12}
\end{align}
 and
\begin{align}
B_{\tiny\mbox{CS}}^{\mu\nu\sigma}\left(p,k;x,z\right) & =i m \widetilde{B}_{\tiny\mbox{CS}}^{\mu\nu\sigma}\left(p,k;x,z\right)+2im^{3}\epsilon^{\mu\nu\sigma},\label{eq: 1.13}
\end{align}
where we have defined the quantity
\begin{align}
\widetilde{B}_{\tiny\mbox{CS}}^{\mu\nu\sigma}\left(p,k;x,z\right) & =2\left(p^{\mu}k_{\alpha}-s^{\mu}s_{\alpha}-2s^{\mu}k_{\alpha}\right)\epsilon^{\alpha\nu\sigma}
+2\left(s_{\alpha}s^{\nu}-2s^{\nu}p_{\alpha}-p_{\alpha}k^{\nu}\right)\epsilon^{\alpha\mu\sigma}\nonumber \\
 & -2\left(s_{\alpha}s^{\sigma}+2p_{\alpha}k^{\sigma}\right)\epsilon^{\alpha\mu\nu}+2p_{\alpha}k_{\beta}\eta^{\nu\sigma}\epsilon^{\alpha\mu\beta}. \label{eq: 1.14}
\end{align}
Hence, replacing the expressions \eqref{eq: 1.12} and \eqref{eq: 1.13} back into \eqref{eq: 1.5}, we find the following expression for the Chern-Simons contribution:
\begin{align}
\Xi_{\tiny\mbox{CS}}^{\mu\nu\sigma}\left(p,k\right) & =\frac{1}{2\pi}sgn\left(m\right)\int d\xi\left[\epsilon^{\mu\nu\sigma}+\frac{1}{8m^{2}}\bigg(6\epsilon^{\mu\nu\sigma}A^{2}
\left(p,k\right)+\widetilde{B}_{\tiny\mbox{CS}}^{\mu\nu\sigma}\left(p,k;x,z\right)\bigg)\right].\label{eq: 1.15}
\end{align}
From the expression \eqref{eq: 1.15} we see that, differently from the contributions appearing in the $AA$-term, \eqref{eq: 0.13a}, we now have HD terms with different indices in the antisymmetric Levi-Civit\`{a} tensor, this shows that these HD terms arising from $\widetilde{B}_{\tiny\mbox{CS}}^{\mu\nu\sigma}$ contribute in a more involving form in terms of derivatives than those arising from $\epsilon^{\mu\nu\sigma}A^{2}$.
Nonetheless, the remaining integrals in \eqref{eq: 1.15} can be readily evaluated to give
\begin{align}
\int d\xi A^{2}\left(p,k\right) & =\frac{1}{12}\left(p^{2}+\left(p.k\right)+k^{2}\right),\label{eq: 1.16}
\end{align}
and
\begin{align}
\int d\xi\widetilde{B}_{\tiny\mbox{CS}}^{\mu\nu\sigma} & =\frac{1}{6}\bigg[\epsilon^{\alpha\nu\sigma}(\frac{5}{2}p^{\mu}k_{\alpha}-p^{\mu}p_{\alpha}+
\frac{1}{2}k^{\mu}p_{\alpha}+3k^{\mu}k_{\alpha})
  +\epsilon^{\alpha\mu\sigma}(-3p_{\alpha}p^{\nu}-k_{\alpha}p^{\nu}
 +k_{\alpha}k^{\nu}-3p_{\alpha}k^{\nu})\nonumber \\
  & -\epsilon^{\alpha\mu\nu}(p_{\alpha}p^{\sigma}
  -k_{\alpha}p^{\sigma}+k_{\alpha}k^{\sigma}+5p_{\alpha}k^{\sigma}
 )\bigg]+\epsilon^{\alpha\mu\beta}\eta^{\nu\sigma}
 p_{\alpha}k_{\beta}.\label{eq: 1.17}
\end{align}
Finally, to obtain the second part of the Chern-Simons and its HD extension contributions to the effective action,
we should replace the above results in such a way: first \eqref{eq: 1.16} and \eqref{eq: 1.17} into
\eqref{eq: 1.15}, and then substituting the resulting expression back into \eqref{eq: 0.3} (for the case $n=3$), with the choice $p_{1}=p$ and $p_{2}=k$ then $p_{3}=-p-k$,
after such substitution we get
\begin{align}
i\Gamma_{\tiny\mbox{CS}}[AAA] & =\frac{g^{3}}{3} \int d^{3}x_{1}d^{3}x_{2}d^{3}x_{3}~A_{\mu}(x_{1})A_{\nu}(x_{2})A_{\sigma}(x_{3})\notag \\
 & \times\int\frac{d^{3}p}{(2\pi)^{3}}\frac{d^{3}k}{(2\pi)^{3}}\exp[-ip.(x_{1}-x_{3})-ik.(x_{2}-x_{3})]\exp[-\frac{i}{2}(p\times k)]~\Xi_{\tiny\mbox{CS}}^{\mu\nu\sigma}. \label{eq: 1.18}
\end{align}
Thus, after some manipulation involving the momentum factors becoming derivatives, and integrations involving the noncommutativity between the gauge fields, we are able to find the final expression for the Chern-Simons effective action for the $AAA$-gauge fields,
\begin{align}
i\Gamma_{\tiny\mbox{CS}}[AAA] & =\frac{g^{3}}{12\pi}~sgn(m)\int d^{3}x~\varepsilon^{\mu\nu\sigma}A_{\mu}\star A_{\nu}\star A_{\sigma}\nonumber \\
 & -\frac{g^{3}}{48\pi m^{2}}~sgn(m)\int d^{3}x~\bigg\{\frac{3}{4}~\varepsilon^{\mu\nu\sigma}\Box A_{\mu}\star A_{\nu}\star A_{\sigma} +\varepsilon^{\alpha\mu\beta}\eta^{\nu\sigma}\partial_{\alpha}A_{\mu}\star\partial_{\beta}A_{\nu}\star A_{\sigma}\nonumber\\
 &   +\frac{1}{12}\varepsilon^{\alpha\nu\sigma}\bigg[13\partial_{\alpha}A_{\sigma}\star\partial^{\mu}A_{\nu}\star A_{\mu}-8\partial^{\mu}A_{\nu}\star\partial_{\alpha}A_{\mu}\star A_{\sigma}-3\partial^{\mu}A_{\nu}\star\partial_{\alpha}A_{\sigma}\star A_{\mu}
  \nonumber\\&+6\partial_{\alpha}A_{\mu}\star\partial^{\mu}A_{\nu}\star A_{\sigma}\bigg]\bigg\}. \label{eq: 1.19z}
\end{align}
The first term in \eqref{eq: 1.19z} is the known $AAA$-part of the NC Chern-Simons action, while the remaining terms are those HD contributions. Moreover, as aforementioned, the HD contributions of the $AAA$-fields, \eqref{eq: 1.19z}, are more involving than those from the $AA$-term, \eqref{eq: 0.13a}. These HD terms give new types of derivative interactions.
The extended Chern-Simons action, the sum of Eqs.\eqref{eq: 0.15} and \eqref{eq: 1.19z}, can be cast conveniently into the following form
\begin{align}
i\Gamma_{\tiny\mbox{CS}}=i\Gamma_{\tiny\mbox{CS}}^{^{(0)}}+i\Gamma_{\tiny\mbox{CS}}^{^{\tiny\mbox{HD}} },
\end{align}
where we have defined, with ${\cal A}_{\mu}=-ig A_{\mu}$, the NC Chern-Simons action
\begin{align}
i\Gamma_{\tiny\mbox{CS}}^{^{(0)}}= \frac{i}{8\pi}sgn(m)\varepsilon^{\mu\nu\sigma}\int d^{3}x~\bigg[
{\cal A}_{\mu}\left(x\right)\partial_{\nu}{\cal A}_{\sigma}\left(x\right)+\frac{2}{3}
{\cal A}_{\mu}(x)\star {\cal A}_{\nu}(x)\star {\cal A}_{\sigma}(x)\bigg], \label{eq: 1.19a}
 \end{align}
and the higher-derivative extended NC Chern-Simons action
\begin{align}
 i\Gamma_{\tiny\mbox{CS}}^{^{\tiny\mbox{HD}}}&=\frac{i}{96\pi m^{2}}sgn(m)\int d^{3}x~\bigg\{ \epsilon^{\mu\nu\sigma} {\cal A}_{\mu}(x)~\square\partial_{\nu}{\cal A}_{\sigma}\left(x\right)
 + \frac{3}{4}~\varepsilon^{\mu\nu\sigma}\Box A_{\mu}\star A_{\nu}\star A_{\sigma} \nonumber\\&+\varepsilon^{\alpha\mu\beta}\eta^{\nu\sigma}\partial_{\alpha}A_{\mu}\star\partial_{\beta}A_{\nu}\star A_{\sigma}
   +\frac{1}{12}\varepsilon^{\alpha\nu\sigma}\bigg[13\partial_{\alpha}A_{\sigma}\star\partial^{\mu}A_{\nu}\star A_{\mu}-8\partial^{\mu}A_{\nu}\star\partial_{\alpha}A_{\mu}\star A_{\sigma}\nonumber\\&-3\partial^{\mu}A_{\nu}\star\partial_{\alpha}A_{\sigma}\star A_{\mu}
  +6\partial_{\alpha}A_{\mu}\star\partial^{\mu}A_{\nu}\star A_{\sigma}\bigg]\bigg\}. \label{eq: 1.19b}
 \end{align}
  By means of dimensional analysis, it can be shown that there are other
 terms contributing to  the higher-derivative Chern-Simons action at this order, arising from the
 AAAA and the AAAAA part (see appendix \ref{sec:appA}).
 Hence, obviously, the relevant effective action would not be gauge invariant,
 without considering these parity violating terms. However, the NC Chern-Simons action $\Gamma_{\tiny\mbox{CS}}^{^{(0)}}$ in \eqref{eq: 1.19a} is explicitly gauge invariant under the infinitesimal gauge transformation $\delta {\cal A} _\mu  = \partial _\mu \lambda + \left[{\cal A} _\mu , \lambda\right]_\star +\mathcal{O} (\lambda^2)$.
At last, just by means of complementarity, we can evaluate the Maxwell contributions from \eqref{eq: 1.10} and \eqref{eq: 1.11},
\begin{align}
A_{\tiny\mbox{M}}^{\mu\nu\sigma}\left(p,k;x,z\right) & =2\left(\left(x-1\right)p+\left(1-z\right)k\right)^{\sigma}\eta^{\mu\nu}
+2\left(\left(x-1\right)p-\left(1+z\right)k\right)^{\mu}\eta^{\nu\sigma}\notag \\
&+2\left(\left(1+x\right)p+\left(1-z\right)k\right)^{\nu}\eta^{\mu\sigma}, \label{eq: 1.20}
\end{align}
 and
\begin{align}
B_{\tiny\mbox{M}}^{\mu\nu\sigma}\left(p,k;x,z\right) & =2m^{2}\left(\left(1-x\right)p+\left(1+z\right)k\right)^{\mu}\eta^{\nu\sigma}-2m^{2}\left(\left(1+x\right)p+\left(1-z\right)k\right)^{\nu}\eta^{\mu\sigma}\notag \\
&+2m^{2}\left(\left(1-x\right)p+\left(z-1\right)k\right)^{\sigma}\eta^{\mu\nu}. \label{eq: 1.21}
\end{align}
Finally, replacing these results back into \eqref{eq: 1.15}, we obtain the $AAA$-fields part for the NC Maxwell action,
\begin{align}
\Xi_{\tiny\mbox{M}}^{\mu\nu\sigma}\left(p,k\right) & =\frac{i}{12\pi}\frac{1}{\left|m\right|}\left[\left(-p+k\right)^{\sigma}\eta^{\mu\nu}-\left(p+2k\right)^{\mu}\eta^{\nu\sigma}+\left(2p+k\right)^{\nu}\eta^{\mu\sigma}\right].\label{eq: 1.22}
\end{align}
\begin{figure}[t]
\vspace{-2.5cm}
  \includegraphics[width=12cm,height=8cm]{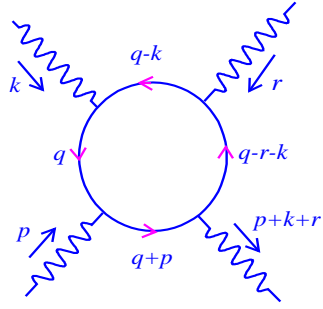}
   \vspace{-1.2cm}\centering
  \caption{Relevant graph for the AAAA-term.
}\label{AAAA-graph}
\end{figure}
\noindent
\subsection{AAAA-term contribution}
By means of complementarity, let us present here the resulting expression for the one-loop contribution \eqref{eq: 0.5} for the case $n=4$, which is shown in Fig.~\ref{AAAA-graph}. This contribution is important in a way to provide
the remaining $AAAA$-fields piece for the NC Maxwell action. The general expression to be evaluated is
\begin{equation}
\Xi^{\mu\nu\sigma\rho}\left(p,k,r\right)=\int\frac{d^{3}q}{\left(2\pi\right)^{3}}
\frac{tr\left[\left(\displaystyle{\not}q+\displaystyle{\not}p+m\right)\gamma^{\mu}
\left(\displaystyle{\not}q+m\right)\gamma^{\nu}\left(\displaystyle{\not}q
-\displaystyle{\not}k+m\right)
\gamma^{\sigma}\left(\displaystyle{\not}q-\displaystyle{\not}r-\displaystyle{\not}k+m\right)
\gamma^{\rho}\right]}{\left[\left(q+p\right)^{2}-m^{2}\right]\left[q^{2}-m^{2}\right]
\left[\left(q-k\right)^{2}-m^{2}\right]\left[\left(q-r-k\right)^{2}-m^{2}\right]}.\label{eq: 2.1}
\end{equation}
Its analysis follows by the same guidelines as presented above for the previous contributions. Hence, in order to obtain the term of interest from the $AAAA$-contribution, let us consider an expansion $m^{2}\gg M^{2}\left(p,k,r\right)$ [in which $M^{2}\left(p,k,r\right)$ is a function of the external momenta]; this provides the following contributions at order $\mathcal{O}\left(m^{-1}\right)$,
\begin{align}
\Xi_{\tiny\mbox{M} }^{\mu\nu\sigma\rho}\left(p,k,r\right) & =\frac{i}{12\pi}\frac{1}{\left|m\right|}\biggl[\eta^{\mu\nu}\eta^{\sigma\rho}
-\eta^{\mu\sigma}\eta^{\nu\rho}+\eta^{\mu\rho}\eta^{\nu\sigma}\biggr]. \label{eq: 2.2}
\end{align}
Therefore, by adding the pieces \eqref{eq: 0.16}, \eqref{eq: 1.22}, and \eqref{eq: 2.2} we find the complete expression of the NC Maxwell action, $\Gamma_{\tiny\mbox{M}}\sim \frac{1}{\left|m\right|} \int d^3 x~ {\cal F} _{\mu\nu} \star {\cal F} ^{\mu\nu}$, in which the field strength tensor has the following expression: ${\cal F} _{\mu\nu}=\partial _{\mu}{\cal A} _{\nu} - \partial _{\nu}{\cal A} _{\mu}+  \left[{\cal A} _{\mu} , {\cal A} _{\nu} \right]_{\star}$.
\section{Propagating modes} \label{sec:4}
In order to characterize the excitations described by the HD contribution as derived in the previous section, let us consider the following higher-order derivative modification
into the pure NC Chern-Simons Lagrangian density and analyze the induced corrections to the propagator pole.
In particular, we shall analyze the infrared sector of the theory, where we expect that the noncommutative UV/IR effects may be present and turn gauge theory unstable. By simplicity we shall consider the following free Lagrangian:
\begin{equation}
\mathcal{L}=\frac{m}{2}\epsilon^{\mu\nu\lambda}A_{\mu}\left(1+\frac{\square}{M^{2}}\right)\partial_{\nu}A_{\lambda}-\frac{1}{2\xi}\left(\partial^{\mu}A_{\mu}\right)^{2},\label{eq: 5.1}
\end{equation}
where we assume $m$ and $M$ both positive. Now, by taking $M=m$, the propagator in the Landau gauge, $\xi=0$, is given by
\begin{equation}
iD_{\mu\nu}\left(p\right)=m\frac{i\epsilon_{\mu\nu\lambda}p^{\lambda}}{p^{2}\left(p^{2}-m^{2}\right)}.\label{eq: 5.2}
\end{equation}
In addition to the free part Eq.\eqref{eq: 5.1}, we shall take the following interacting part \cite{ref77},\footnote{Although this Lagrangian does not constitute the whole contribution,
we have considered some of the parts obtained from the derivative contributions for the vertex functions
that suffice to our interest in a quantitative discussion for the NC and HD contributions in the dispersion relation; the remaining terms would contribute as changing the numerical factors in the resulting outcome, but this does not change our quantitative analysis.}
\begin{align}
\mathcal{L}_{int} & =mg\varepsilon^{\mu\nu\sigma}\left(\frac{2}{3}A_{\mu}\star A_{\nu}\star A_{\sigma}+\frac{1}{6m^{2}}\square A_{\mu}\star A_{\nu}\star A_{\sigma}\right) \nonumber \\
& +\frac{mg} {12m^{2}}\varepsilon^{\alpha\beta\gamma}\eta^{\lambda\delta}\partial_{\alpha}A_{\beta}\star\partial_{\gamma}A_{\lambda}\star A_{\delta}+\frac{img^{2}}{2m^{2}}\varepsilon^{\mu\nu\sigma}A_{\mu}\star A_{\nu}\star\partial^{\rho}A_{\sigma}\star A_{\rho}, \label{eq: 5.3}
\end{align}
besides, we also have the ghost term,
\begin{equation}
\mathcal{L}_{gh}= \partial^{\mu}\overline{c}\star D_{\mu}c,\label{eq: 5.4}
\end{equation}
with $D_{\mu}c= \partial _\mu c - ig \left[A_\mu , c\right]_\star$. We are here interested in evaluating the one-loop photon self-energy diagrams. The respective Feynman rules can be obtained from the above Lagrangian densities, the $AAA$, $AAAA$ and $\bar{c}Ac$ vertex functions.\footnote{Notice that the $AAAAA$ vertex, which is needed to ensure the gauge invariance of the HD parts (see appendix A), does
not contribute to the one-loop diagrams.} Actually, the diagrams needed to be computed have the same form as in the usual NC Chern-Simons theory,
however we have now obtained different (derivative) Feynman rules.

For the purpose of calculating the leading terms of the ultraviolet contribution \cite{ref78}, we can set $|p| = 0$, while keeping $\tilde{p}$ on the phase factor in the vertex functions, which can lead to the UV/IR mixing. Finally, after evaluating the three diagrams contributing at one-loop order, we have the complete planar contribution \cite{ref77}
\begin{equation}
\left(\Pi^{\mu\nu}\right)_{p}\left(p\right)=\frac{mg^{2}}{60\pi}\eta^{\mu\nu}, \label{eq: 5.5}
\end{equation}
whereas, the complete non-planar contribution reads
\begin{align}
\left(\Pi^{\mu\nu}\right)_{n-p}\left(p\right) & =-\frac{g^{2}}{36\pi}\biggl\{
\left[\eta^{\mu\nu}-\left(1+\left|\tilde{p}\right|m\right)
\frac{\tilde{p}^{\mu}\tilde{p}^{\nu}}{\tilde{p}^{2}}\right]
\frac{e^{-\left|\tilde{p}\right|m}}{\left|\tilde{p}\right|}+6m^{2}\int_{0}^{1}dx\left(1-x\right)
\left[\frac{1}{\alpha}\eta^{\mu\nu}-\left|\tilde{p}\right|\frac{\tilde{p}^{\mu}\tilde{p}^{\nu}}{\tilde{p}^{2}}\right]
e^{-\left|\tilde{p}\right|\alpha}\nonumber \\
 & +9m^{4}\int_{0}^{1}dx\int_{0}^{1-x}dy\left(1-x-y\right)
 \left[\frac{1}{\beta^{3}}\left(1+\left|\tilde{p}\right|\beta\right)
 \eta^{\mu\nu}-\frac{1}{\beta}\tilde{p}^{\mu}\tilde{p}^{\nu}\right]e^{-\left|\tilde{p}\right|\beta}\biggr\}\nonumber \\
 & +\frac{g^{2}}{4\pi}\frac{1}{\left|\tilde{p}\right|}\left[\eta^{\mu\nu}-\frac{\tilde{p}^{\mu}\tilde{p}^{\nu}}
 {\tilde{p}^{2}}\right]-\frac{g^{2}}{4\pi}\frac{1}{\left|\tilde{p}\right|}\int_{0}^{1}dx
 \left[\eta^{\mu\nu}-\left(1+\left|\tilde{p}\right|\alpha\right)\frac{\tilde{p}^{\mu}
 \tilde{p}^{\nu}}{\tilde{p}^{2}}\right]e^{-\left|\tilde{p}\right|\alpha}, \label{eq: 5.6}
\end{align}
where $\alpha^{2}=\left(1-x\right)m^{2}$ and $\beta^{2}=\left(1-x-y\right)m^{2}$.

In order to discuss the radiative correction effects into the photon propagator pole, it is interesting to recall that due to gauge symmetry constrain the tensor structure to be
\begin{equation}
\Pi_{\mu\nu}\left(p\right)=\left(\eta_{\mu\nu}-\frac{p_{\mu}p_{\nu}}{p^{2}}\right)
\Pi_{\tiny\mbox{S}}\left(p^{2}\right)+\frac{\tilde{p}_{\mu}
\tilde{p}_{\nu}}{\tilde{p}^{2}}\Pi_{\tiny\mbox{NC}}\left(p^{2}\right)+i\epsilon_{\mu\nu\lambda}p^{\lambda}
\Pi_{\tiny\mbox{A}}\left(p^{2}\right). \label{eq: 5.7}
\end{equation}
Moreover, we can determine the expression for the complete propagator by making use of the following identity: $\left(\mathcal{D}^{-1}\right)^{\mu\nu}\left(p\right)  =\Gamma^{\mu\nu}\left(p\right)-\Pi^{\mu\nu}\left(p\right)$ and that $\left(\mathcal{D}^{-1}\right)^{\mu\nu}\mathcal{D}_{\nu\lambda}=i\delta_{\lambda}^{\mu}$.
After some algebraic manipulation we obtain that
\begin{equation}
i\mathcal{D}_{\mu\nu}=\frac{ \Pi_{S}+ \Pi_{NC}}{\cal R}\left(\eta_{\mu\nu}-\frac{p_{\mu}p_{\nu}}{p^{2}}\right)
+\frac{\xi}{p^{2}}\frac{p_{\mu}p_{\nu}}{p^{2}}-\frac{ \Pi_{NC}}{\cal{R}}\frac{\tilde{p}_{\mu}\tilde{p}_{\nu}}{\tilde{p}^{2}}-\frac{\frac{1}{m}\left(p^{2}-m^{2}\right)+
\Pi_{A}}{\cal{R}}i\epsilon_{\mu\nu\lambda}p^{\lambda},\label{eq: 5.8}
\end{equation}
where ${\cal{R}}=\Pi_{\tiny\mbox{S}}\left(\Pi_{\tiny\mbox{S}}+ \Pi_{\tiny\mbox{NC}}\right)-p^{2}\left(\frac{1}{m}\left(p^{2}-m^{2}\right)+\Pi_{\tiny\mbox{A}}\right)^{2}$.
From the above results, at one-loop order, we immediately see that $\Pi_{\tiny\mbox{A}}=0$. Now, the remaining form factors, in the leading contributions when $\left|\tilde{p}\right|\rightarrow 0$,
\begin{align}
\Pi_{\tiny\mbox{NC}}\left(p^{2}\right) & =\frac{g^{2}}{36\pi}\frac{1}{\left|\tilde{p}\right|}+\mathcal{O}\left(m\left|\tilde{p}\right|\right);\quad\Pi_{\tiny\mbox{S}}\left(p^{2}\right)=-\frac{7g^{2}}{20\pi}m-\frac{g^{2}}{24\pi}\frac{1}{\left|\tilde{p}\right|}+\mathcal{O}\left(g^{2}\left|\tilde{p}\right|\right). \label{eq: 5.9}
\end{align}
We notice that the last term of \eqref{eq: 5.8} can be rewritten as $ \frac{m}{\left(p^{2}-m^{2}\right)}
\frac{i\epsilon_{\mu\nu\lambda}p^{\lambda}}{p^{2}-\frac{m^{2}}{\left(p^{2}-m^{2}\right)^{2}}\Pi_{\tiny\mbox{S}}
\left(\Pi_{\tiny\mbox{S}}+\Pi_{\tiny\mbox{NC}}\right)}$. \\
\noindent
This shows that the HD pole
remains intact and only the massless mode, $p^2=0$, receives contribution from the radiative corrections.
From this we obtain the dispersion relation,
\begin{equation}
\omega^{2}\left(\mathbf{p}\right)=\mathbf{p}^{2}+\frac{1}{m^{2}}\Pi_{\tiny\mbox{S}}\left(\Pi_{\tiny\mbox{S}}
+\Pi_{\tiny\mbox{NC}}\right)+\mathcal{O}\left(g^{2}\frac{\left|p\right|}{m^{2}}\right). \label{eq: 5.10}
\end{equation}
Now, the tree-level parameter can be conveniently rewritten $g^{2}\sim m/\kappa$,
we finally find at low momenta,
\begin{align}
\omega^{2}\left(\mathbf{p}\right) & =\mathbf{p}^{2}+m_{eff}^{2}+\frac{1}{1728\pi^{2}}\frac{m^{4}}{\kappa^{2}}\frac{h^{2}}
{\mathbf{p}^{2}}+\frac{7}{360\pi^{2}}\frac{m^{3}}{\kappa^{2}}\frac{h}{\left|\mathbf{p}\right|}+
\mathcal{O}\left(\frac{\left|p\right|}{m\kappa}\right)+\mathcal{O}\left(\frac{m\left|\tilde{p}\right|}
{\kappa}\right), \label{eq: 5.11}
\end{align}
where we have defined $m_{eff}^{2}=\frac{49}{400\pi^{2}}\frac{m^{2}}{\kappa^{2}}$ and $h^{-1}=m^{2}\theta$. In contrast with the analysis in Ref.~\cite{ref79},
we have no instability in the infrared when $\left|\mathbf{p}\right|/m\ll h$.
There the authors have considered a different expression for $\Pi_{S}$,
where no $1 / \left|\tilde{p}\right|$ term is present; this consideration has changed the sign of
the $h$ term in \eqref{eq: 5.11} to negative, implying therefore that the range of applicability of the above expression was for $\left|\mathbf{p}\right|/m\ll h$. Moreover, this consideration has
also excluded the most infrared singular $1 / \left|\tilde{p}\right|^2$ contribution.
\section{Concluding remarks}
\label{sec:5}
In this paper we have studied the one-loop effective action of NC QED$_3$, fermionic fields interacting with an Abelian gauge field. After integrating out the fermionic fields, we obtained the effective action for the noncommutative gauge field, next we have computed explicitly the contributions of the $n=2,3,4$ terms for the effective action. In addition, it should be remarked that the noncommutative effects into the resulting outcome are in fact nonperturbative. This fact is supported, in a formal way, in the existence of an exact Seiberg-Witten map, valid to all orders in $\theta$.

In completion to previous analyses, we have supplied with detailed calculation of the $n=2,3,4$ terms for the effective action. Based in the exact calculation, we have considered, in particular, the long wavelength limit (large fermion mass) into the terms in the expansion $\mathcal{O}(m^ 0)$, $\mathcal{O}(m^ {-1})$ and $\mathcal{O}(m^{-2})$, that corresponded exactly to the NC Chern-Simons action, NC Maxwell action, and higher-derivative extension to the NC Chern-Simons action, respectively. Moreover, the gauge invariance of the higher-derivative extension was also discussed. \\
\noindent
In addition, we introduced a new Lagrangian density including the pure NC Chern-Simons term with its higher-derivative extension. Then, the one-loop effect of the higher-derivative terms to the pole of the photon propagator and the relevant dispersion relation in the infrared limit was addressed.

Certainly a study considering the effects of these higher-derivative terms in the gauge fields into the theory's quantities would be rather interesting \cite{ref27}, as well as a detailed study about the finite gauge invariance of these higher-derivative NC gauge fields \cite{ref20,ref28}. Those aspects are currently under scrutiny.
\subsection*{Acknowledgments}
The authors would like to thank the anonymous referee for his/her comments and
suggestions to improve this paper. M.Gh. is grateful to M.M. Sheikh-Jabbari for fruitful discussions and also thanks
School of Physics of Institute for research in fundamental sciences (IPM) for the partial support and for the research facilities and environment. R.B. acknowledges FAPESP for full support, Project No. 2013/26571-4.

\appendix

\section{Fixing the general structure of the HD-terms}
\label{sec:appA}
In this appendix, we use the dimensional analysis to discuss carefully the gauge invariance of the higher-derivative extended NC Chern-Simons action $\Gamma_{\tiny\mbox{CS}}^{^{\tiny\mbox{HD}}}$ in \eqref{eq: 1.19b}.
The mass dimension of the gauge and fermionic fields in $d$ dimensions is given by
\begin{eqnarray}
[A]=\frac{d-2}{2},\quad [\psi]=\frac{d-1}{2}.
\end{eqnarray}
For the case $d=3$, we have $[A]=\frac{1}{2}$ and $[\psi]=1$. Hence the mass dimension of the coupling constant $g$ (minimal coupling) would be $[g]=\frac{1}{2}$. For the sake of our discussion it is useful to introduce ${\cal A}_{\mu}=-ig A_{\mu}$, which is concluded $[{\cal A}]=1$.

\noindent
Now, considering the ordinary CS term: ${\cal A}\partial{\cal A}$ with $[{\cal A}\partial{\cal A}]=3$, that other possible contribution with mass dimension of 3 is ${\cal A}{\cal A}{\cal A}$ (non-Abelian or noncommutative self-interacting gauge fields). Hence, by dimensional analysis, the CS action without higher-derivative (HD) terms is given by a linear combination of ${\cal A}\partial{\cal A}$ and ${\cal A}{\cal A}{\cal A}$.

\noindent
What about the HD terms? Suppose that we know the structure of the HD terms appearing in the AA-part, hence, we then try to guess recursively the general structure of HD terms needed to be added into the AA-part in order to make a gauge invariant higher-derivative action. The mass dimension of the simplest HD contribution in the AA-term is given by $[{\cal A}\partial\Box{\cal A}]=5$, so we can make all of the possible contributions using ${\cal A}$ and $\partial$ with total mass dimension of 5. It is important to emphasize that due to the Levi-Civit\`{a} tensor $\epsilon ^{\mu \nu \lambda}$ and metric $g^{\rho\sigma}$, the several terms are produced just like those appearing in relation \eqref{eq: 1.19b} at the order of $sgn(m)/m^2$:
\begin{align}
&\Gamma^{(2)}:~~{\cal A}\partial\Box{\cal A} , \label{eq: a.1}\\
&\Gamma^{(3)}:~~{\cal A}\partial{\cal A}\partial{\cal A},\quad \Box{\cal A}{\cal A}{\cal A} ,\label{eq: a.2}\\
&\Gamma^{(4)}:~~\partial{\cal A}{\cal A}{\cal A}{\cal A} , \label{eq: a.3}\\
&\Gamma^{(5)}:~~{\cal A}{\cal A}{\cal A}{\cal A}{\cal A} \label{eq: a.4}.
\end{align}
We then can conclude that these arguments suffice to determine the general structure of such terms, but only the explicit calculation of their contributions that allows us to determine the tensorial coefficients of these different terms, although the gauge invariance imposes some limitations.

To show the correctness of the above argument, let us consider the dimensional analysis of the loop integrals that we have mentioned, $n=4$ in \eqref{eq: 0.5}, \footnote{For the contributions as \eqref{eq: a.1} and \eqref{eq: a.2} see \eqref{eq: 0.15} and \eqref{eq: 1.19b}, respectively.}
\begin{align}
\Xi^{\mu\nu\rho\sigma}\sim\int\frac{d^{\omega}q}{(2\pi)^{\omega}}~\frac{
\frac{1}{\omega(\omega+2)}~q^{4}C^{\mu\nu\rho\sigma}+\frac{1}{\omega}q^{2}D^{\mu\nu\rho\sigma}
+E^{\mu\nu\rho\sigma}}{\bigg[q^{2}+M^{2}-m^{2}\bigg]^{4}},
\end{align}
in which each contribution labeled by its coefficient in the leading order is given by
\begin{align}
&C:~\int\frac{d^{\omega}q}{(2\pi)^{\omega}}~\frac{q^{4}}{(q^{2}+M^{2}-m^{2})^{4}}
\sim\frac{1}{(M^{2}-m^{2})^{4-2-\frac{\omega}{2}}}\sim
\frac{1}{|m|} , \nonumber\\
&D:~\int\frac{d^{\omega}q}{(2\pi)^{\omega}}~\frac{q^{2}}
{(q^{2}+M^{2}-m^{2})^{4}}\sim\frac{1}{(M^{2}-m^{2})^{4-1-\frac{\omega}{2}}}\sim \frac{1}{|m|}\frac{1}{m^{2}} ,\nonumber\\
&E:~\int\frac{d^{\omega}q}{(2\pi)^{\omega}}~\frac{1}
{(q^{2}+M^{2}-m^{2})^{4}}\sim\frac{1}{(M^{2}-m^{2})^{4-\frac{\omega}{2}}}\sim \frac{1}{|m|}\frac{1}{m^{4}} .
\end{align}
By means of dimensional analysis, it is deduced that
\begin{itemize}
\item There is not any power of $m$ in $C^{\mu\nu\rho\sigma}$.

\item  There is an $m$ term in $D^{\mu\nu\rho\sigma}$ and $m^3$ in $E^{\mu\nu\rho\sigma}$ that contribute to our result, which are corresponding to those terms with one-derivative structure: $\partial{\cal A}{\cal A}{\cal A}{\cal A}$ in \eqref{eq: a.3}.
\end{itemize}
 Now, by completeness, we study the case $n=5$ of \eqref{eq: 0.5}:
\begin{align}
\Xi^{\mu\nu\rho\sigma\lambda}\sim\int\frac{d^{\omega}q}{(2\pi)^{\omega}}~\frac{
\frac{1}{\omega(\omega+2)}~q^{4}C^{\mu\nu\rho\sigma\lambda}+\frac{1}{\omega}q^{2}D^{\mu\nu\rho\sigma\lambda}
+E^{\mu\nu\rho\sigma\lambda}}{\bigg[q^{2}+M^{2}-m^{2}\bigg]^{5}},
\end{align}
where
\begin{align}
&C:~\int\frac{d^{\omega}q}{(2\pi)^{\omega}}~\frac{q^{4}}
{(q^{2}+M^{2}-m^{2})^{5}}\sim\frac{1}{(M^{2}-m^{2})^{5-2-\frac{\omega}{2}}}\sim \frac{1}{|m|}\frac{1}{m^{2}},\nonumber\\
&D:~\int\frac{d^{\omega}q}{(2\pi)^{\omega}}~ \frac{q^{2}}
{(q^{2}+M^{2}-m^{2})^{5}}\sim\frac{1}{(M^{2}-m^{2})^{5-1-\frac{\omega}{2}}} \sim \frac{1}{|m|}\frac{1}{m^{4}}, \nonumber\\
&E:~\int\frac{d^{\omega}q}{(2\pi)^{\omega}}~\frac{1}
{(q^{2}+M^{2}-m^{2})^{5}}\sim\frac{1}{(M^{2}-m^{2})^{5-\frac{\omega}{2}}}\sim \frac{1}{|m|}\frac{1}{m^{6}}.
\end{align}
We can then see that
\begin{itemize}
  \item  There is an $m$ term in $C^{\mu\nu\rho\sigma}$, an $m^3$ in $D^{\mu\nu\rho\sigma}$, and $m^5$ in $E^{\mu\nu\rho\sigma}$ that contribute to our result. These contributions lead to those terms without any derivative: $~{\cal A}{\cal A}{\cal A}{\cal A}{\cal A}$ in \eqref{eq: a.4}.
\end{itemize}
Here, we would like to generalize this analysis for the generic higher-derivative terms that may appear in the next orders.
Since we have expanded our result in powers of $(\frac{\Box}{m^{2}})$ in \eqref{eq: 0.13}, in general, we can call the order of the expansion $\ell$ and start from the AA-part,
\begin{align}
{\cal A}\partial(\frac{\Box}{m^{2}})^{\ell}{\cal A}=\frac{1}{m^{2\ell}}{\cal A}\partial\Box^{\ell}{\cal A},
\end{align}
in which the mass dimension of ${\cal A}\partial\Box^{\ell}{\cal A}$ is given by
\begin{align}
[{\cal A}\partial\Box^{\ell}{\cal A}]=3+2\ell .
\end{align}
Therefore, we can determine the general structure of the different higher-derivative expressions with the mass dimension of $3+2\ell$ using combinations of ${\cal A}$ and $\partial$. Actually, these expressions contribute to the gauge invariant NC Chern-Simons action $\Gamma^{t}_{\ell}$ at order $\ell$.
Inserting the several values for $\ell$, we obtain
\begin{itemize}
  \item $\ell=0~\mathbf{\rightarrow}~[\Gamma^{t}_{\ell}]=3~\mathbf{\rightarrow}
      ~\Gamma^{t}_{0}=\alpha_{0}\Gamma^{(2)}_{0}+\beta_{0}\Gamma^{(3)}_{0}$.
\end{itemize}
Here, $t$ indicates the total action. We note that the case $\ell=0$ corresponds to the ordinary noncommutative Chern-Simons action with $\alpha_{0}=1$ and $\beta_{0}=\frac{2}{3}$:
\begin{align}
\Gamma^{(2)}_{0}\sim {\cal A}\partial{\cal A}, \quad \Gamma^{(3)}_{0}\sim {\cal A}{\cal A}{\cal A}.
\end{align}
\begin{itemize}
\item $\ell=1~\mathbf{\rightarrow}~[\Gamma^{t}_{\ell}]=5~\mathbf{\rightarrow}~\Gamma^{t}_{1}=
    \alpha_{1}\Gamma^{(2)}_{1}+\beta_{1}\Gamma^{(3)}_{1}+\rho_{1}\Gamma^{(4)}_{1}+
    \sigma_{1}\Gamma^{(5)}_{1}$,
\end{itemize}
in which
\begin{align}
&\Gamma^{(2)}_{1}\sim {\cal A}\partial\Box{\cal A}; \quad \Gamma^{(3)}_{1}\sim
\Box{\cal A}{\cal A}{\cal A},~\partial{\cal A}\partial{\cal A}{\cal A};\quad \Gamma^{(4)}_{1}\sim\partial{\cal A}{\cal A}{\cal A}{\cal A}; \quad  \Gamma^{(5)}_{1}\sim{\cal A}{\cal A}{\cal A}{\cal A}{\cal A}.
\end{align}
\begin{itemize}
\item $\ell=2~\mathbf{\rightarrow}~[\Gamma^{t}_{\ell}]=7~\mathbf{\rightarrow}~\Gamma^{t}_{2}=
    \alpha_{2}\Gamma^{(2)}_{2}+\beta_{2}\Gamma^{(3)}_{2}+
    \rho_{2}\Gamma^{(4)}_{2}+\sigma_{2}\Gamma^{(5)}_{2}+\lambda_{2}\Gamma^{(6)}_{2}+\xi_{2}\Gamma^{(7)}_{2}$
\end{itemize}
in which
\begin{align}
&\Gamma^{(2)}_{2}\sim {\cal A}\partial\Box^{2}{\cal A} ; \quad \Gamma^{(3)}_{2}\sim \Box^{2}{\cal A}{\cal A}{\cal A},\quad \partial{\cal A}\partial{\cal A}\Box{\cal A}, \quad
\partial\Box{\cal A}\partial{\cal A}{\cal A}, \quad \Box{\cal A}\Box{\cal A}{\cal A} ; \\
&\Gamma^{(4)}_{2}\sim\partial\Box{\cal A}{\cal A}{\cal A}{\cal A}, \quad
\partial{\cal A}\Box{\cal A}{\cal A}{\cal A}, \quad \partial{\cal A}\partial{\cal A}\partial{\cal A}{\cal A}; \\
&\Gamma^{(5)}_{2}\sim\Box{\cal A}{\cal A}{\cal A}{\cal A}{\cal A}, \quad \partial{\cal A}\partial{\cal A}{\cal A}{\cal A}{\cal A}; \quad
\Gamma^{(6)}_{2}\sim\partial{\cal A}{\cal A}{\cal A}{\cal A}{\cal A}{\cal A}; \quad
\Gamma^{(7)}_{2}\sim{\cal A}{\cal A}{\cal A}{\cal A}{\cal A}{\cal A}{\cal A} .
\end{align}
 All of the coefficients appearing in $\Gamma^{t}_{\ell}$ are determined using the explicit calculations such that the gauge invariance of the total NC Chern-Simons action is preserved.


\end{document}